\documentclass[aps, twocolumn,showpacs,amsmath,amssymb,floatfix,amstex]{revtex4-1}

\usepackage{amsfonts}
\usepackage{graphicx}
\newcommand{\ket}[1]{| #1\rangle}

\begin{document}
\title{Multifractal dimensions for all moments for certain critical random matrix ensembles in the strong multifractality regime}
\author{E. Bogomolny and O. Giraud}
\affiliation{Univ.~Paris-Sud, CNRS, LPTMS, UMR8626, F-91405, Orsay, France}

\date{December 9, 2011}

\begin{abstract}
We construct perturbation series for the $q$th moment of eigenfunctions of various critical random matrix ensembles in the strong multifractality regime close to localization. Contrary to previous investigations, our results are valid in the region $q<1/2$. Our findings allow to verify, at first leading orders in the strong multifractality limit, the symmetry relation for anomalous fractal dimensions $\Delta_q=\Delta_{1-q}$, recently conjectured for critical models where an analogue of the metal-insulator transition takes place. It is known that this relation is verified at leading order in the weak multifractality regime. Our results thus indicate that this symmetry holds in both limits of small and large coupling constant. For general values of the coupling constant we present careful numerical verifications of this symmetry relation for different critical random matrix ensembles. We also present an example of a system closely related to one of these critical ensembles, but where the symmetry relation, at least numerically, is not fulfilled.  
\end{abstract}

\maketitle

\section{Introduction}
It was realized more than fifty years ago by Anderson \cite{anderson} that disorder can lead to localization of wave functions in quantum systems. Depending on the strength of disorder, three-dimensional wave functions at a given energy are either all delocalized (metallic phase) or all localized (insulator phase).  In recent years, such Anderson transitions have been observed experimentally in various media: localization of Bose-Einstein condensates into disorder created by laser speckle \cite{bec}, localization of ultrasound in elastic media \cite{ultrasound}, localization of cold atoms in a kicked potential \cite{delande}. 

Many works have investigated theoretically the properties of quantum disordered systems exactly at the metal-insulator transition \cite{evers}. A characteristic feature of wave functions at the transition point is that they are neither localized nor delocalized. Their localization properties are described by a set of multifractal exponents characterizing the behavior of the wave function moments at different scales. More precisely, if we consider a wave function $\ket{\Psi}$ belonging to an $N$-dimensional Hilbert space with components $\Psi_j$, then the $q$th moment of the wave function behaves for large $N$ as
$\sum_{j=1}^N |\Psi_j|^{2q}\propto N^{-D_q(q-1)}$. If  $\ket{\Psi}$ is localized then $D_q=0$ for all positive $q$, while if the state is delocalized over the whole Hilbert space then $D_q$ is the dimension of the system itself. Multifractality is characterized by the fact that $D_q$ is a a nontrivial function of $q$. 

Understanding the properties of multifractal dimensions has been the subject of intense research in the past few years (see e.g.~\cite{evers} and references therein). In order to investigate the mechanisms involved, various models have been proposed. The simplest models where an analogue of the metal-insulator transition can be observed and studied are random Hermitian or unitary matrices whose off-diagonal elements $M_{{\bf m}{\bf n}}$ decrease as a power of the distance from the diagonal, as
\begin{equation}
M_{{\bf m}{\bf n}}\sim \frac{g}{|{\bf m}-{\bf n}|^r},\,\,\,\, |{\bf m}-{\bf n}|\gg 1,
\label{property}
\end{equation}
where $g$ plays the role of a coupling constant (here ${\bf m}$, ${\bf n}$ are $d$--dimensional vectors, where $d$ is the space dimension). When $r>d$, eigenvectors of these matrices are localized; they are delocalized for $r<d$. Anderson-like transition occurs in the localization properties of the eigenvectors at $r=d$, and typically a multifractal behavior is observed \cite{levitov}, similar to  the Anderson model at the point of  metal-insulator transition \cite{mit}. Such ensembles for $r=d$ are often called critical random matrix ensembles. The main attention in this field was given to the investigation of the ensemble of so-called critical banded random matrices (CrBRM) \cite{seligman}, whose matrix elements $M_{mn}$ are independently distributed Gaussian variables with zero mean and variance decreasing away from the main diagonal as in \eqref{property} with $r=d=1$. Other models with similar properties, such as ultrametric critical random matrix ensembles, have also been proposed \cite{ossipov}. Recently other critical random matrix ensembles, based on Lax matrices of classical one-dimensional $N$--particle integrable systems, were introduced in \cite{lax}. It was shown that the spectrum of such matrices is of intermediate type \cite{lax2}, and that the corresponding eigenvectors display multifractal properties \cite{bogomolny_giraud}. 

Most of the analytical results for these critical ensembles are obtained within perturbation series expansions in the regime of strong multifractality where eigenvectors are almost localized, or in the regime of weak multifractality where eigenvectors are almost entirely extended. Nevertheless, a few relations \cite{universality, entropy} are believed to be universal and valid for a large (but not well defined) class of critical ensembles. One of the most striking ones involves anomalous fractal dimensions $\Delta_q$ defined by
\begin{equation}
\Delta_q=(D_q-d)(q-1)\ ,
\end{equation}
which characterize the departure of multifractal exponents from their value in the delocalized case (since $\Delta_q=0$ for extended states). In \cite{universality}  it was conjectured that these anomalous dimensions have to be symmetric with respect to the axis $q=1/2$, namely
\begin{equation}
\Delta_q=\Delta_{1-q}.
\label{symmetry}
\end{equation}
One of the analytical arguments in favor of this  result is the four-loop calculation of $\Delta_q$ within nonlinear $\sigma$-model in $2+\varepsilon$ dimensions \cite{wegner}, which manifests symmetry \eqref{symmetry}. Recently, it has been argued that this relation holds exactly in models which can be mapped onto nonlinear $\sigma$-models \cite{mirlin11}. This is for instance the case for critical random matrix models in the regime of weak multifractality. The fact that this symmetry is not always verified is illustrated by the existence of critical systems for which such a symmetry relation does not hold in general \cite{simplemaps} (see also Section \ref{sec_interm}). Thus the domain of validity of \eqref{symmetry} remains unclear.
 
For general systems, the perturbation expansion approach remains the main analytical tool for investigating multifractal dimensions. For systems of the form \eqref{property}, analytical expressions for multifractal dimensions at leading orders of perturbation expansion have been obtained in the domain $q>1/2$, both in the strong and weak multifractality regimes \cite{bogomolny_giraud}. While the methods used to derive these results are still applicable in the domain $q<1/2$ in the weak multifractality regime, no perturbation expansion approach has been developed (to the best of authors' knowledge) in the domain $q<1/2$ in the strong multifractality case. Therefore direct analytical verification of the symmetry  \eqref{symmetry} in the regime of small $g$ was not yet possible. 

The goal of this paper is to derive analytical expressions for the $D_q$ in the domain $q<1/2$ where previous approaches did break down. We obtain a perturbative expression for $q<1/2$ in the regime of strong multifractality, and perform extensive numerical calculations which corroborate our analytical findings. This enables us to test analytically the correctness of the symmetry relation \eqref{symmetry} for several different critical random matrix ensembles. The main conclusion of the paper is that, at least at leading orders in the perturbation series, the symmetry is fulfilled.

The outline of the paper is the following. In section \ref{sec_models} the models investigated are introduced. Then multifractal dimensions $D_q$ are defined in section \ref{sec_definition}. The perturbation expansion of $D_q$ in the strong multifractality regime for $q<1/2$ yields a nontrivial result even at lowest order; the corresponding calculation is done in section \ref{sec_zero}. In section \ref{sec_first} the calculation is performed at the next-to-leading order. This enables to demonstrate the validity of the symmetry relation \eqref{symmetry} up to this order for strong multifractality. Together with known results for the weak multifractality regime, this proves that the symmetry holds in both extreme regimes. In order to assess the validity of this relation in general, numerical results are presented for the different models considered in section \ref{sec_figs}. In section \ref{sec_interm} we present numerical results for a system without perturbation theory expansion, where eigenfunctions display nontrivial multifractal dimensions, but for which the symmetry property \eqref{symmetry} is not valid. In section \ref{sec_toy}, a toy model related to the above ensembles, and where exact analytical calculation of multifractal dimensions is possible, is briefly discussed.

\section{The models}\label{sec_models}
We now introduce the models that we consider in this paper. Four critical random matrix ensembles associated with the Lax matrices of classical integrable systems have been introduced in \cite{bogomolny_giraud}. Calogero-Moser (CM) $N$--particle systems yield three ensembles of $N\times N$ Hermitian matrices of the form
\begin{equation}
M_{mn}=p_m\delta_{mn}+\mathrm{i} g(1-\delta_{mn})V(m-n), 
\label{CM}
\end{equation}
where $p_m$ are independent Gaussian random variables with zero mean and unit variance, and $V(k)$ is one of the three following functions
\begin{equation} 
 \frac{1}{k},\qquad \frac{\mu}{N\sinh(\mu k/N)}, \qquad \frac{\mu}{N\sin(\mu k/N)}. 
\end{equation}
with $\mu$ a parameter ($\delta_{mn}$ is the Kronecker symbol). We denote these ensembles respectively by CM$_r$, CM$_h$ and CM$_t$. The fourth ensemble introduced in \cite{bogomolny_giraud} corresponds to the Lax matrix of Ruijsenaars-Schneider (RS) system. It consists of a set of unitary $N\times N$ matrices which, up to a simple transformation, reduce to
\begin{equation}
M_{mn}=\mathrm{e}^{\mathrm{i}\Phi_m}\frac{\sin\pi g }{N\sin\left(\pi(m-n+g\right)/N)},
\label{RS}
\end{equation}
where $\Phi_m$ are independent random phases uniformly distributed between $0$ and $2\pi$ and $g$ is some fixed constant. A closely related model, introduced in \cite{giraud} and investigated in \cite{Schmit}, corresponds to a family of quantum maps obtained by quantization of a classical map on the torus. These quantum maps have exactly the form \eqref{RS}, but with $g$ now depending on the matrix size, namely $g=aN$ with $a$ fixed. Spectral statistics of these ensembles have been shown to be of intermediate type \cite{BogDubSch}.

We will also consider the well-studied CrBRM ensemble, which is the set of $N\times N$ symmetric or Hermitian matrices whose elements $M_{mn}$ are independently distributed Gaussian variables with zero mean and variance
\begin{eqnarray}
\langle |M_{nn}|^2\rangle&=&\frac{1}{\beta},\nonumber\\
\langle |M_{mn}|^2 \rangle &=&\frac{1}{2}\left [1+\Big (\frac{m-n}{g}\Big )^2 \right ]^{-1},\,\,\, m\neq n.
\label{CrBRME}
\end{eqnarray}
Here $\beta=1$ if the matrices considered are real symmetric, or $\beta=2$ if they are complex Hermitian.

\section{Multifractal dimensions and symmetry relation}\label{sec_definition}
We want to investigate multifractality properties of the above ensembles of random matrices. Multifractality manifests itself in a non-trivial scaling of eigenfunction moments. Let $\Psi_n(\alpha)$ be a normalized eigenfunction of the matrix $M_{mn}$ corresponding to eigenvalue $\lambda_{\alpha}$, so that
\begin{equation}
\sum_{n=1}^N M_{mn}\Psi_{n}(\alpha)=\lambda_{\alpha} \Psi_{m}(\alpha).
\label{eq_M}
\end{equation}
Mean eigenfunction moments are defined for any real $q$ as $I_q=\langle \sum_j|\Psi_j|^{2q}\rangle$, where the average is taken over all random variables of the model and over a small window of eigenvalues. More precisely, we set
\begin{equation}
I_q=\frac{1}{N}\sum_{\alpha=1}^N\frac{1}{\rho(E)}\Big \langle \sum_{j=1}^N|\Psi_j(\alpha)|^{2q}\delta(E-\lambda_{\alpha}) \Big \rangle \ ,
\label{I_q}
\end{equation}
where $\rho(E)$ is the total mean eigenvalue density
\begin{equation}
\label{rhoE}
\rho(E)=\frac{1}{N}\Big \langle \sum_{\alpha=1}^N \delta(E-\lambda_{\alpha})\Big \rangle . 
\end{equation} 
Multifractal dimensions $D_q$ are defined as the scaling exponents in the large $N$ behavior of these moments,
\begin{equation}
I_q \underset{N\to \infty}{\sim} N^{-D_q(q-1)}\ .
\label{D_q}
\end{equation}
In our matrix models, the  dimension $d$ of the underlying space is 1, so that anomalous dimensions are given by $\Delta_q=(D_q-1)(q-1)$. Both analytical and numerical calculation of these exponents is a difficult task. Indeed, it is so far not possible to have direct access to exact analytical expressions, and numerical calculations require diagonalization of large matrices. There is however a way to get some insight into analytical expressions by constructing perturbation series for the eigenfunction moments for small or large values of the coupling constant \cite{evers, levitov, kravtsov_2}. For moments of order $q$ with $q>1/2$, the lowest order of perturbation series at small and large $g$ has a universal dependence on $q$ for all critical ensembles considered above. In the regime of weak multifractality, the unperturbed eigenvectors are delocalized, and at first order
 \begin{equation}
D_q\simeq 1-t q, 
\label{dqweak}
\end{equation} 
where the constant $t$ depends on the system. It is equal to $1/(2\pi\beta g)$ for CrBRME, and 0 for CM models, where this regime corresponds to $g\to\infty$. For RS ensemble, weak multifractality regime is reached whenever $g$ is close to a nonzero integer, and  $t=(g-k)^2/k^2$, where $k$ is the integer closest to $g$ \cite{bogomolny_giraud}. Equation \eqref{dqweak} is valid for $t |q|\ll 1$. In the regime of strong multifractality ($|g|\ll 1$), the unperturbed matrices (corresponding to $g=0$) are diagonal and their eigenvectors are localized. The first-order correction is given by
\begin{equation}
D_q\underset{g\to 0}{\simeq}  4| g| \rho(E)\, s\, \frac{ \sqrt{\pi}\, \Gamma \Big (q-\frac{1}{2}\Big)}{ \Gamma(q)}\ ,
\label{dqstrong}
\end{equation}
again valid only for $q>1/2$. Here $\rho(E)$ is the density of states of unperturbed system, which, at that order, is equal to the actual density of states \eqref{rhoE}. The constant $s$ in \eqref{dqstrong} can be obtained as the coefficient in front of the logarithmic term
\begin{equation}
\Big \langle \frac{1}{N}\sum_{\overset{m,n=1}{m\neq n}}^N | M_{mn}| \Big \rangle\underset{N\to\infty}{\sim} 2s\ln N \ .  
\label{Sfirst_order}
\end{equation}
The explicit expressions of $s$ can be found e.g.~in \cite{bogomolny_giraud}. It is equal to 1 for RS, CM$_r$ and CM$_h$ ensembles, $1/\sqrt{\pi}$ for symmetric CrBRME, $\sqrt{\pi/8}$ for Hermitian CrBRME, and $[\mu/\pi]$ for CM$_t$, where $[.]$ denotes the integer part. 
 
In the weak multifractality case at leading order, the fact that $D_q$ is linear in $q$ (Eq.~\eqref{dqweak}) implies that the symmetry relation holds. Indeed, $\Delta_q\simeq -q(q-1)t$ is automatically symmetric with respect to the substitution $q\to 1-q$. In \cite{rushkin} it was demonstrated analytically  that for CrBRME and ultrametric ensembles the second order term also obeys this symmetry. In the same  paper the third order term has been calculated numerically for these models and it has been found that it also fulfilled \eqref{symmetry}. Results in \cite{mirlin11} suggest that these results extend to all orders of perturbation theory in the weak multifractality regime.

The situation in the strong multifractality limit is less clear, since \eqref{dqstrong} is valid only when $q>1/2$ \cite{evers}. The purpose of the present paper is to obtain an expression for $D_q$ in this regime when $q<1/2$. We show that it can be derived within the standard perturbation series approach, and yields, for $q<1/2$, 
\begin{equation}
D_q \underset{g\to 0}{\simeq} \frac{2q-1}{q-1}+4 |g| \rho(E)\, s\, \frac{ \sqrt{\pi}\, \Gamma \Big (\frac{1}{2}-q\Big)}{(q-1)\Gamma(-q)}\ .
\label{dqstrong_sym}
\end{equation}
The existence  of a $g$-independent term in \eqref{dqstrong_sym} and its absence in \eqref{dqstrong} when $q>1/2$ looks, at first glance, as the necessity of non-analytical contributions. In fact, as we shall see, the  perturbation series is obtained in an even simpler way than the one used to derive  \eqref{dqstrong}. 

A remarkable feature of \eqref{dqstrong_sym} is that it coincides with what would be obtained from \eqref{dqstrong} assuming that the symmetry \eqref{symmetry} holds. Therefore this perturbation theory approach shows that the symmetry \eqref{symmetry} is correct at order zero and at order 1 in the coupling constant $g$ for all ensembles considered.

\section{Zero-order term}\label{sec_zero}

Let us consider a set of random Hermitian matrices of the form
\begin{equation}
M_{mn}=p_m\delta_{mn} +g(1-\delta_{mn})\mu_{mn}, 
\label{matrix_simple}
\end{equation}
where $p_m$ are independent identically distributed random variables distributed according to some probability distribution $\sigma(p)$. The off-diagonal elements $\mu_{mn}$ may contain random variables, but all are independent on the $p_m$. Matrices for CM and CrBRM ensemble are obviously of that form. The case of the RS ensemble will be discussed separately in section \ref{sectionRS}.

Standard perturbation theory \cite{morse} applied to matrix \eqref{matrix_simple} for $g\ll 1$ yields eigenvalues
\begin{equation}
\label{lambdaseries}
\lambda_{\alpha}=p_{\alpha}+g^2\sum_{n\neq\alpha}\frac{|\mu_{n\alpha}|^2}{p_{\alpha}-p_n}+\mathcal{O}(g^3)
\end{equation}
 and eigenfunctions
\begin{equation}
\Psi_m(\alpha)=g\frac{\mu_{m\alpha}}{p_{\alpha}-p_m}+g^2\sum_{ n\neq m, \alpha}\frac{\mu_{mn}\mu_{n\alpha}}{(p_{\alpha}-p_n)(p_{\alpha}-p_m)}+\mathcal{O}(g^3)
\label{series}
\end{equation}
for $m\neq \alpha$, and $\Psi_{\alpha}(\alpha)=1+\mathcal{O}(g^2)$. 
In the first non-vanishing order, for fixed $\alpha$, we neglect the term in $g^2$ in \eqref{series} and get
\begin{eqnarray}
\label{firstorder}
\Big \langle \sum_{\genfrac{}{}{0pt}{}{m=1}{m\neq \alpha}}^N|\Psi_m(\alpha)|^{2q}\delta(E-\lambda_{\alpha}) \Big \rangle =\hspace{2cm}\\
\nonumber
\hspace{1cm}g^{2q}\sum_{m\neq \alpha}\langle|\mu_{m\alpha} |^{2q}\rangle\Big \langle\frac{\delta(E-p_{\alpha})}{(p_{\alpha}-p_m)^{2q}}\Big \rangle\ ,
\end{eqnarray}
since by assumption the $p_j$ are independent on the $\mu_{mn}$ and $\lambda_{\alpha}=p_{\alpha}+\mathcal{O}(g^2)$. In order to calculate the averages one has to integrate over random variables $p_n$ distributed according to the law $\sigma(p)$. The average over variables $p_\alpha$ and $p_m$ with $m\neq \alpha$ reads
\begin{eqnarray}
\label{zero_order_mean}
\Big \langle\frac{\delta(E-p_{\alpha})}{(p_{\alpha}-p_m)^{2q}}\Big \rangle&=&\int\mathrm{d}p_{\alpha}\mathrm{d}p_{m}\sigma(p_{\alpha})\sigma(p_{m})\frac{\delta(E-p_{\alpha})}{(p_{\alpha}-p_m)^{2q}}\nonumber\\
&=&\sigma(E)A_q,
\end{eqnarray}
where we have set
\begin{equation}
\label{integralAq}
A_q=\int\mathrm{d}p\frac{\sigma(p)}{(E-p)^{2q}}.
\end{equation}
This integral is finite whenever $q<1/2$. From Eqs.~\eqref{rhoE} and \eqref{lambdaseries} one has 
\begin{equation}
\rho(E)=\sigma(E)+\mathcal{O}(g^2),
\end{equation}
thus the moment \eqref{I_q} reads
\begin{equation}
I_q=1+\frac{g^{2q}A_q}{N}\sum_{\alpha}\sum_{m\neq\alpha}\big \langle |\mu_{m\alpha} |^{2q} \big \rangle .
\label{sum_0}
\end{equation}
The summation in \eqref{sum_0} can be performed in the same way as was done e.g.~in \cite{bogomolny_giraud} to obtain \eqref{Sfirst_order}, but instead of the $\ln N$ behavior of \eqref{Sfirst_order}, for $q<1/2$ these quantities will scale as $N^{1-2q}$ when $N\to\infty$. For instance in the case of CM$_r$ model, $\mu_{mn}=\mathrm{i}/(m-n)$, and for $q<1/2$
\begin{equation} 
\label{term_0}
\frac{1}{N} \sum_{\underset{m\neq n}{m,n}}\frac{1}{|m-n|^{2q}}
=\frac{2}{N} \sum_{k=1}^{N-1}\frac{N-k}{k^{2q}}
\underset{N\to \infty}{\sim}\frac{N^{1-2q}}{(1-2q)(1-q)},
\end{equation}
so that for large $N$, the mean moment $I_q$ scales as
\begin{equation}
I_q \underset{N\to \infty}{\sim}g^{2q}\frac{A_q}{(1-2q)(1-q)}N^{1-2q}.
\label{asymptotic}
\end{equation}
In the trigonometric case, one has
\begin{eqnarray}
\frac{1}{N} \sum_{\underset{m\neq n}{m,n}}\frac{\mu^{2q}}{N^{2q}\sin^{2q}(\mu(m-n)/N)}\hspace{2cm}\\
\hspace{2cm}\underset{N\to \infty}{\sim}2\mu^{2q}N^{1-2q}\int_0^1\mathrm{d}y\frac{1-y}{\sin^{2q}(\mu y)},\nonumber
\end{eqnarray}
so that 
\begin{equation}
I_q\underset{N\to \infty}{\sim}g^{2q}\mu^{2q}A_q\int_0^1\mathrm{d}y\frac{2(1-y)}{\sin^{2q}(\mu y)}N^{1-2q}.
\label{asymptotic_sin}
\end{equation}
For the other ensembles considered, the calculation is similar (technical details are the same as the ones encountered for calculating  \eqref{Sfirst_order}, see \cite{bogomolny_giraud}). All these ensembles yield the same $N^{1-2q}$ behavior. 

The definition \eqref{D_q} of fractal dimensions then allows us to extract the leading order term for multifractal dimensions in the domain $q<1/2$ as
\begin{equation}
D_q=\frac{2q-1}{q-1}.
\label{dqzero}
\end{equation}
Note that for $q>1/2$, the various integrals appearing in the above calculation do not converge. Thus, this usual perturbation theory approach breaks down. Different techniques were used in the case $q>1/2$, yielding an expression for $D_q$ given by Eq.~\eqref{dqstrong}. At order zero this expression \eqref{dqstrong} gives $D_q=0$. In particular, together with Eq.~\eqref{dqzero}, these results show that the symmetry relation \eqref{symmetry} is fulfilled at order zero of perturbation theory.

\begin{figure}[ht]
\begin{center}
\includegraphics[width=.95\linewidth]{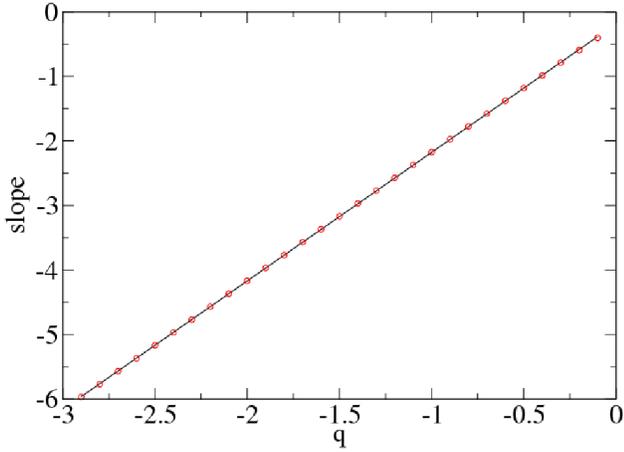}
\end{center}
\caption{(Color online) Circles: Slopes $a$ obtained by a linear fit $\ln(I_q/N^{1-2q})=a\ln g +b$, versus $q$, for CM$_t$ ensemble. The slope is extracted from values of $g$ from $5. 10^{-5}$ to $2. 10^{-3}$. Matrix size is $N=2^{10}+1$, average is taken over 64 central eigenvectors (i.e.~around $E=0$) and 1600 realizations. Solid line is the best linear fit $a = -0.186 + 1.99 q$ (to be compared with the expected $a=2q+$constant from \eqref{asymptotic_sin}). \label{fig1}}
\end{figure}

In order to assess the correctness of the $g^{2q}$ behavior in the leading-order term in the expressions \eqref{asymptotic}--\eqref{asymptotic_sin}, we take as an example the case of the trigonometric Calogero-Moser model in Fig.~\ref{fig1}. According to \eqref{asymptotic_sin}, the ratio $I_q/N^{1-2q}$ is expected to behave as $C_q g^{2q}$ for small values of $g$, with $C_q$ a factor depending on $q$. In logarithmic scale we thus expect straight lines with slopes $2q$. The comparison with this predicted power, given in Fig.~\ref{fig1}, is very good. 

\section{First-order term}\label{sec_first}

To get the first-order correction to the above result one has to take into account the term in $g^2$ in \eqref{series}. The modulus square of the wave function with $m\neq \alpha$ is then given by
\begin{eqnarray}
\label{second_term}
|\Psi_m(\alpha)|^2&=&\frac{g^2}{(p_{\alpha}-p_m)^2}\Big(|\mu_{m\alpha}|^2\\
&+&g\sum_{n\neq m,\alpha}\frac{\mu_{mn}\mu_{n\alpha}\mu_{m\alpha}^*+\textrm{c.c.}}{p_{\alpha}-p_n}\nonumber\\
&+&g^2\sum_{n,n'\neq m,\alpha}\frac{\mu_{mn}\mu_{n\alpha}\mu_{mn'}^*\mu_{n'\alpha}^*}{(p_{\alpha}-p_n)(p_{\alpha}-p_{n'})}\Big)\nonumber
\end{eqnarray} 
(c.c.~denotes the complex conjugate). As for all critical systems, the dominant correction in such formulae when $g\to 0$ corresponds to transitions between two states $n$ and $\alpha$ with a small energy difference $p_n-p_{\alpha}$ of the order of $g$ \cite{levitov}. Taking more states in resonance into account gives higher-order corrections \cite{levitov}. Calculation of the first-order term depends on whether the off-diagonal matrix elements are pure imaginary or not, since in the pure imaginary case the term in $g^3$ vanishes in \eqref{second_term}. Since matrices for CM ensembles have this feature, we first consider this case.

\subsection{CM ensembles}
CM ensembles defined by \eqref{CM} have the property that their off-diagonal elements are pure imaginary numbers. In this case \eqref{second_term} gives
\begin{eqnarray}
\label{PsimCM}
|\Psi_m(\alpha)|^{2q}=\frac{g^{2q}|\mu_{m\alpha}|^{2q}}{(p_{\alpha}-p_m)^{2q}}\hspace{4cm}\\
\hspace{1cm}\times\Big(1+\frac{g^2}{|\mu_{m\alpha}|^2}\sum_{n,n'\neq m,\alpha}\frac{\mu_{mn}\mu_{n\alpha}\mu_{mn'}^*\mu_{n'\alpha}^*}{(p_{\alpha}-p_n)(p_{\alpha}-p_{n'})}\Big)^q\ .\nonumber
\end{eqnarray} 
Note that at this order it is still possible to replace $\sigma(E)$ by $\rho(E)$ and $\delta(E-\lambda_{\alpha})$ by $\delta(E-p_{\alpha})$ since corrections are of higher order. The average $\langle\delta(E-p_{\alpha})/(p_{\alpha}-p_m)^{2q}\rangle$ over variables $p_n$ in \eqref{firstorder} is now replaced by
\begin{equation}
\Big\langle\frac{\delta(E-p_{\alpha})}{(p_{\alpha}-p_m)^{2q}}
\Big(1+\frac{g^2}{|\mu_{m\alpha}|^2}\hspace{-.3cm}\sum_{n,n'\neq m,\alpha}\frac{\mu_{mn}\mu_{n\alpha}\mu_{mn'}^*\mu_{n'\alpha}^*}{(p_{\alpha}-p_n)(p_{\alpha}-p_{n'})}\Big)^q\Big\rangle\ .
\label{second_term_only}
\end{equation}
Since the variables $p_n$ and $p_{n'}$ appearing in \eqref{second_term_only} are different from the variables $p_{\alpha}$ and $p_m$, the average \eqref{second_term_only} can be done separately over variables $p_n, p_{n'}$. For the term involving these variables, the correction to the order zero is given by 
\begin{equation}
\Big\langle
\Big(1+\frac{g^2}{|\mu_{m\alpha}|^2}\sum_{n,n'\neq m,\alpha}\frac{\mu_{mn}\mu_{n\alpha}\mu_{mn'}^*\mu_{n'\alpha}^*}{(p_{\alpha}-p_n)(p_{\alpha}-p_{n'})}\Big)^q-1\Big\rangle\ ,
\label{second_term_correction}
\end{equation}
where the average is performed only over variables $p_n$  with $n\neq m,\alpha$. For fixed $g$, the main contribution to \eqref{second_term_correction}, upon integration over random variables, will come from terms with $p_n\simeq p_{\alpha}$. The leading-order correction \eqref{second_term_correction}  can in fact be replaced by 
\begin{equation}
\sum_{n\neq m, \alpha}
\Big\langle\Big (1+\frac{g^2}{|\mu_{m\alpha}|^2} \frac{|\mu_{mn}\mu_{n\alpha}|^2}{(p_{\alpha}-p_n)^2}\Big )^q-1\Big\rangle\ .
\label{second_term_only_diagonal}
\end{equation}
In order to show this, one has to check that the difference between \eqref{second_term_correction} and \eqref{second_term_only_diagonal} is of higher order in $g$. This can be done by upper bounding the integrand of the difference (in the integral over $p_n$) uniformly in $g$ by a function whose integral converges at $g=0$. Note that a similar problem occurs in the calculation of the zero-order term \eqref{zero_order_mean} which we obtained in the previous section: for fixed $g$ the average \eqref{second_term_only} diverges at $p_n\simeq p_{\alpha}$, but nevertheless \eqref{second_term_correction} converges to 0 when $g\to 0$, which proves that \eqref{zero_order_mean} gives the correct zero-order contribution.

The term \eqref{second_term_only_diagonal} is a sum over $n$. For a given fixed index $n$, the contribution is
\begin{equation}
\label{contribution_pn}
J_n(g)=\int\mathrm{d}p_n\sigma(p_n)\left[\Big (1+\frac{g^2}{|\mu_{m\alpha}|^2} \frac{|\mu_{mn}\mu_{n\alpha}|^2}{(p_{\alpha}-p_n)^2}\Big )^q-1\right]\ .
\end{equation} 
Changing variables by the transformation 
\begin{equation}
p_n=p_\alpha+g\frac{|\mu_{mn}\mu_{n\alpha}|}{|\mu_{m\alpha}|}t\ ,
\end{equation}
and keeping only lowest-order terms in $g$ in the expansion of $\sigma(p_n)$, the contribution \eqref{contribution_pn} gives
\begin{equation}
\label{Jng}
J_n(g)=|g|\frac{|\mu_{mn}\mu_{n\alpha}|}{|\mu_{m\alpha}|}\sigma(p_{\alpha})\int_{-\infty}^{\infty}  \left [ \left(1+\frac{1}{t^2}\right)^q-1\right] \mathrm{d}t\ .
\end{equation}
The integral can be calculated for $q<1/2$ (see e.g.~\cite{bateman}), yielding
\begin{equation}
\label{bq}
B_q=\int_{-\infty}^{\infty}  \left [ \Big (1+\frac{1}{t^2}\Big )^q-1\right ] \mathrm{d}t = -\frac{\sqrt{\pi}\,\Gamma(1/2-q)}{\Gamma(-q)}\ .
\end{equation}
The sum of all contributions in \eqref{second_term_only_diagonal} gives the first-order correction term
\begin{eqnarray}
\label{earlier}
|g|\Big\langle\frac{\delta(E-p_{\alpha})}{(p_{\alpha}-p_m)^{2q}}\sum_{n\neq m, \alpha}
\frac{|\mu_{mn}\mu_{n\alpha}|}{|\mu_{m\alpha}|}\sigma(p_{\alpha}) B_q\Big\rangle\hspace{1cm}\\
\nonumber
\hspace{1cm}=|g|\rho(E)^2A_qB_q\sum_{n\neq m, \alpha}
\frac{|\mu_{mn}\mu_{n\alpha}|}{|\mu_{m\alpha}|}\ ,
\end{eqnarray}
using the result of \eqref{zero_order_mean}. Putting together equations \eqref{I_q}, \eqref{firstorder}, \eqref{second_term_only} and \eqref{earlier}, the first-order correction to moments $I_q$ is
\begin{equation}
\delta I_q=g^{2q}|g|\frac{\rho(E)A_qB_q}{N}
\sum_{\genfrac{}{}{0pt}{}{m,\alpha}{m\neq\alpha}}\sum_{n\neq m, \alpha}|\mu_{m\alpha}|^{2q}\frac{|\mu_{mn}\mu_{n\alpha}|}{|\mu_{m\alpha}|}\ .
\end{equation}
In the case of CM$_r$ model where $\mu_{mn}=\mathrm{i}/(m-n)$, we need the asymptotic large-$N$ behavior of 
\begin{equation}
\label{sum_first_order}
\frac{1}{N}\sum_{\genfrac{}{}{0pt}{}{m,\alpha}{m\neq\alpha}}\frac{1}{(m-\alpha)^{2q}}\sum_{n\neq m, \alpha}\left|\frac{1}{m-n}-\frac{1}{\alpha-n}\right|\ .
\end{equation}
This sum is dominated by two regions $n=m+k$ with $1\ll k\ll m$ and  $n=\alpha+k$ with $1\ll k\ll \alpha$. In both regions the sum over $k$ diverges logarithmically, so that
\begin{eqnarray}
\sum_{n\neq m, \alpha}\left|\frac{1}{m-n}-\frac{1}{\alpha-n}\right|&\simeq&
\sum_{n\neq m}\frac{1}{|m-n|}+\sum_{n\neq \alpha}\frac{1}{|\alpha-n|}\nonumber\\
&\simeq& 2\ln N\ .
\label{sum_k}
\end{eqnarray}
Using \eqref{term_0} and \eqref{sum_k} for these two regions, we conclude that the behavior of the sum \eqref{sum_first_order} is $\sim 4N^{1-2q}\ln N$ for large $N$. As we show in the following sections, for other ensembles similar calculations yield a behavior in  $4 s N^{1-2q}\ln N$, where $s$ is the same as in \eqref{Sfirst_order}. Thus at first order
\begin{equation}
I_q=g^{2q}C_qN^{1-2q}\left(1+4|g|B_q\rho(E) s\ln N \right)\ ,
\end{equation}
where $B_q$ is given by \eqref{bq}. This can be seen as a first-order expansion of $C_q(g) N^{-D_q(g)(q-1)}$ at small values of $g$. We thus get the first-order correction to the fractal dimensions as
\begin{equation}
\label{DqCM}
D_q=\frac{2q-1}{q-1}+4|g|\rho(E) s\frac{\sqrt{\pi}\,\Gamma(1/2-q)}{(q-1)\Gamma(-q)}\ ,
\end{equation}
which coincides with Eq.~\eqref{dqstrong_sym}. As already mentioned, together with the earlier result \eqref{dqstrong}, this confirms the symmetry relation \eqref{symmetry} at first order of perturbation series at small $g$ for models \eqref{CM}.

\subsection{RS ensemble}\label{sectionRS}
For RS ensemble, expanding \eqref{RS} at small $g$ yields, up to second order in $g$,
\begin{equation}
M_{mm}\simeq\mathrm{e}^{\mathrm{i}\Phi_m}\left(1-\frac{\pi^2g^2(N^2-1)}{6N^2}\right) 
\end{equation}
for diagonal elements and
\begin{eqnarray}
\label{M1RS}
M_{mn}\simeq\frac{\pi g}{N}\mathrm{e}^{\mathrm{i}\Phi_m}\left(\frac{1}{\sin\left(\pi(m-n)/N\right)}\right.\hspace{2cm}\\
\nonumber\hspace{2cm}-\left.\frac{\pi g\cos\left(\pi(m-n)/N\right)}{N\sin^2\left(\pi(m-n)/N\right)}\right)\nonumber
\end{eqnarray}
for off-diagonal elements with $m\neq n$. Thus at that order the matrix $M$ takes the form \eqref{matrix_simple} if we set $p_m=M_{mm}$ and $M_{mn}=g\mu_{mn}$. Keeping only terms up to order $g^2$, the expression \eqref{series} now gives
\begin{eqnarray}
\label{psimRS}
\Psi_m(\alpha)=\frac{\pi g}{N}\frac{\mathrm{e}^{\mathrm{i}\Phi_{m}}}{\mathrm{e}^{\mathrm{i}\Phi_{\alpha}}-\mathrm{e}^{\mathrm{i}\Phi_{m}}}
\Bigg [\frac{1}{\sin(\pi(m-\alpha)/N)}\hspace{1cm}\\
-\frac{\pi g\cos\left(\pi(m-\alpha)/N\right)}{N\sin^2\left(\pi(m-\alpha)/N\right)}+\frac{\pi g}{N}\sum_{n\neq m,\alpha}\frac{\mathrm{e}^{\mathrm{i}\Phi_{n}}}{\mathrm{e}^{\mathrm{i}\Phi_{\alpha}}-\mathrm{e}^{\mathrm{i}\Phi_{n}}}\nonumber\\
\times\frac{1}{\sin\left(\pi(m-n)/N\right)\sin\left(\pi(n-\alpha)/N\right)}\Bigg ].\nonumber
\end{eqnarray}
Using the identity for $m\neq\alpha$
\begin{eqnarray}
\sum_{\underset{n\neq m,\alpha}{n=1}}^N\frac{1}{\sin\left(\pi(m-n)/N\right)\sin\left(\pi(n-\alpha)/N\right)}=\hspace{1.5cm}\\
\nonumber
-\frac{2\cos \left(\pi(m-\alpha)/N\right)}{\sin^2 \left(\pi(m-\alpha)/N\right)} 
\end{eqnarray}
and the fact that
\begin{equation}
 \frac{\mathrm{e}^{\mathrm{i}\Phi_{n}}}{\mathrm{e}^{\mathrm{i}\Phi_{\alpha}}-\mathrm{e}^{\mathrm{i}\Phi_{n}}}=\frac{\mathrm{i}}{2}\cot\frac{\Phi_n-\Phi_{\alpha}}{2}-\frac12,
\end{equation}
one can check that two terms cancel in \eqref{psimRS}, so that $\Psi_m(\alpha)$ reduces to
\begin{eqnarray}
\Psi_m(\alpha)=\frac{\pi g}{N}\frac{\mathrm{e}^{\mathrm{i}\Phi_{\alpha}}}{\mathrm{e}^{\mathrm{i}\Phi_{\alpha}}-\mathrm{e}^{\mathrm{i}\Phi_{m}}}
\Bigg [\frac{1}{\sin(\pi(m-\alpha)/N)}\hspace{1cm}\\
+\frac{\pi g}{2N}\sum_{n\neq m,\alpha}\frac{\mathrm{i}\cot((\Phi_n-\Phi_{\alpha})/2)}{\sin\left(\pi(m-n)/N\right)\sin\left(\pi(n-\alpha)/N\right)}\Bigg ].\nonumber
\end{eqnarray}
This has a form similar to \eqref{series}, with a pure imaginary first-order correction term, for leading-order terms where $\Phi_{\alpha}\sim \Phi_{n}$. It can be treated essentially in the same way, yielding the same expression \eqref{DqCM} for $D_q$, with $s=1$.

\subsection{Critical random matrix ensemble} 

In the above derivation we used the fact that for models \eqref{CM} and \eqref{RS} the term in $g^3$ in \eqref{second_term} vanishes.  For CrBRME \eqref{CrBRME} this is not true and this case requires a different treatment. One can consider this ensemble as a particular case of \eqref{matrix_simple} with  the substitution $p_k=M_{kk}$ and (at first order in $g$) $\mu_{mn}$ distributed as independent Gaussian variables with zero mean and variance  
\begin{equation}
\label{variancemu}
\langle |\mu_{mn}|^2\rangle =\frac{1}{(m-n)^2}\ .
\end{equation} 
Since $\mu_{m\alpha}$ and $\mu_{mn}$ in \eqref{series} are independent complex Gaussian random variables, $\Psi_m(\alpha)$ appears as a sum of random variables whose variances are given by \eqref{variancemu}. It is thus itself a complex Gaussian random variable $Z$ of mean 0 and variance 
\begin{equation}
\langle |Z|^2\rangle =\frac{g^{2}}{(p_{\alpha}-p_m)^{2}}\Bigg( \langle|\mu_{m\alpha}|^2\rangle+g^2\hspace{-.2cm}\sum_{n\neq \alpha ,m}\hspace{-.2cm}\frac{\langle |\mu_{mn}|^2\rangle |\mu_{n\alpha}|^2}{(p_{\alpha}-p_k)^2}\Bigg )
\end{equation} 
(at this stage we consider the $\mu_{n\alpha}$ fixed). For $q>-1$, the mean value of the $q$th power of $Z$ can be calculated as 
\begin{equation}
\langle |Z|^{2q}\rangle=\langle |Z|^2\rangle^q  \Gamma(q+1)\ ,
\end{equation}
so that 
\begin{eqnarray}
\langle |\Psi_m(\alpha)|^{2q}\rangle&=&\Gamma(q+1)\frac{g^{2q} \langle|\mu_{m\alpha}|^2\rangle^q}{(p_{\alpha}-p_m)^{2q}}\\
\nonumber&\times&\left(1 +g^2\sum_{n\neq \alpha ,m}\frac{\langle |\mu_{mn}|^2\rangle |\mu_{n\alpha}|^2}{ \langle|\mu_{m\alpha}|^2\rangle(p_{\alpha}-p_n)^2}\right)^q\ .
\end{eqnarray}
The treatment is thus the same as in the previous section (compare with \eqref{PsimCM}), apart from an overall factor $\Gamma(q+1)$ and with the replacement 
\begin{equation}
 |\mu_{m\alpha}|^{2q} \left |\frac{ \mu_{mn}\mu_{n\alpha}}{\mu_{m \alpha}}\right |
\to
 \langle |\mu_{m\alpha}|^{2}\rangle^q \sqrt{\frac{\langle |\mu_{mn}|^2\rangle}{\langle |\mu_{m \alpha}|^2\rangle}}\langle|\mu_{n\alpha}|\rangle\ .
\end{equation} 
In view of \eqref{variancemu} the calculation is exactly the same as for the CM$_r$ ensembles and yields \eqref{DqCM} for $D_q$ with $s=1$. For $q<-1$ the mean value of the $q$th power of $Z$ is infinite, and mean quantities should be normalized by the typical value of $Z$, as was discussed in detail in \cite{mirlin}. Again, the symmetry \eqref{symmetry} is verified at this order of perturbation theory.

\section{Numerical results} \label{sec_figs}
In previous sections we showed that the symmetry relation \eqref{symmetry} holds at lowest orders of perturbation theory for small coupling constant $g$. For large values of the coupling constant (weak multifractality limit), results have been presented in \cite{bogomolny_giraud}. In order to obtain results for general coupling constant (between the two extreme regimes where perturbation theory holds), we have to resort to numerical investigations. Here we present numerical results for the matrix ensembles considered above at various values of the coupling constant. Matrices are diagonalized and moments of the eigenvectors are calculated for different matrix sizes. The large-$N$ behavior of moments is then extracted via a fit of the form $\log\langle \sum_i|\Psi_i|^{2q}\rangle=a+b\log N+c/N$. For negative values of $q$, moments are replaced by a local averaging $\sum_i(\sum_j|\Psi_{4i+j}|^2)^q$ to avoid divergences.

\begin{figure}[ht]
\begin{center}
\includegraphics[width=.95\linewidth]{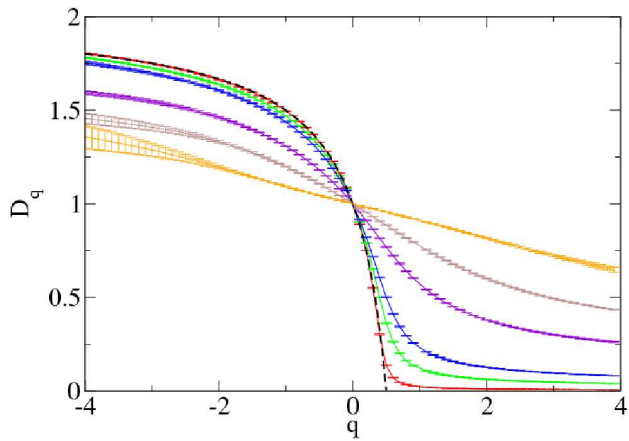}
\end{center}
\caption{(Color online) Fractal dimensions $D_q$ as a function of $q$ for CM$_r$ ensemble for (from bottom to top at $q=4$) $g=0.005$ (red), 0.025 (green), 0.05 (blue), 0.15 (violet), 0.25 (brown), 0.4 (orange). The $p_k$ are independent random variables distributed according to a Gaussian with mean 0 and variance 1. Matrix sizes for numerical fit are $N=2^n$, $8\leq n\leq 12$. Average is performed over the $N/16$ eigenvectors closest to the eigenvalue $E=0$. Number of random realizations of the matrix is between 2560 for $N=2^8$ and 80 for $2^{12}$; error bars correspond to the standard deviation. Dashed curve is the function $q\mapsto (2q-1)/(q-1)$. \label{fig2}}
\end{figure}

Figures \ref{fig2} and \ref{fig4} display the multifractal dimensions $D_q$ as a function of $q$ for CM$_r$ and CM$_t$ ensembles respectively. As CM matrices are Hermitian with a density $\rho(E)$ depending on the energy window considered, the averages are performed only over eigenvectors whose eigenvalues are in the vicinity of $E=0$ (i.e.~in the center of the spectrum). For RS ensemble, multifractal dimensions $D_q$ are shown in Fig.~\ref{fig6}. In this case, matrices are unitary, and averages are performed over all eigenvectors. Figure \ref{fig8} displays multifractal dimensions of CrBRME. For all these models, the plots of $D_q$ follow Eqs.~\eqref{dqstrong} and \eqref{dqstrong_sym} for small values of the coupling constant. This was in fact checked indirectly in \cite{bogomolny_giraud}, where plots of $D_q$ for small $g$ and $q>1/2$ were shown to agree with \eqref{dqstrong}, while plots for $q<1/2$ agreed with the expression obtained from \eqref{dqstrong} assuming the symmetry \eqref{symmetry}. Since this symmetry is now proven analytically in the small $g$ limit, the plots for $q<1/2$ indeed coincide with \eqref{dqstrong_sym}.

\begin{figure}[ht]
\begin{center}
\includegraphics[width=.95\linewidth]{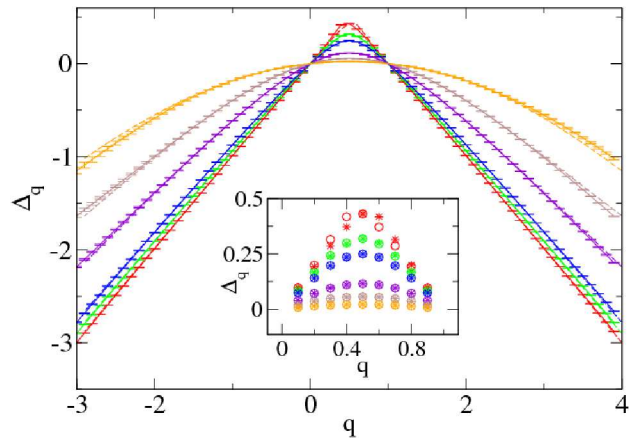}
\end{center}
\caption{(Color online) Anomalous dimensions $\Delta_q=(D_q-1)(q-1)$ (solid) and $\Delta_{1-q}$ (dashed) as a function of $q$ for CM$_r$ ensemble, same data and same color code as in Fig.~\ref{fig2}. $g$ increases from bottom curve to top curve at $q=4$. Inset: same data zoomed in, circles correspond to $\Delta_q$, stars to $\Delta_{1-q}$. For almost all data points stars lie inside circles, which indicates symmetry of $\Delta_q$.\label{fig3}}
\end{figure}

\begin{figure}[ht]
\begin{center}
\includegraphics[width=.95\linewidth]{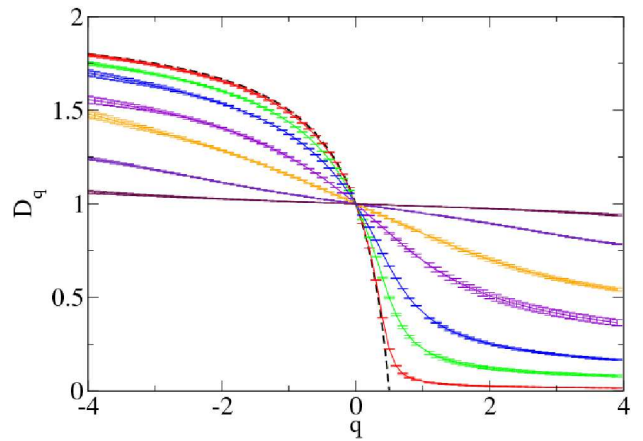}
\end{center}
\caption{(Color online) Fractal dimensions $D_q$ as a function of $q$ for CM$_t$ ensemble for $\mu=2\pi/N$ and (from bottom to top at $q=4$) $g=0.005$ (red), 0.025 (green), 0.05 (blue), 0.1 (violet), 0.15 (orange), 0.25 (indigo), 0.4 (brown). The $p_k$ are independent random variables distributed according to a Gaussian with mean 0 and variance 1. Matrix sizes for numerical fit are $N=2^n$, $8\leq n\leq 12$. Average is performed over the $N/16$ eigenvectors closest to the eigenvalue $E=0$. Number of random realizations of the matrix is between 2560 for $N=2^8$ and 80 for $2^{12}$. Dashed curve is the function $q\mapsto (2q-1)/(q-1)$. \label{fig4}}
\end{figure}

\begin{figure}[ht]
\begin{center}
\includegraphics[width=.95\linewidth]{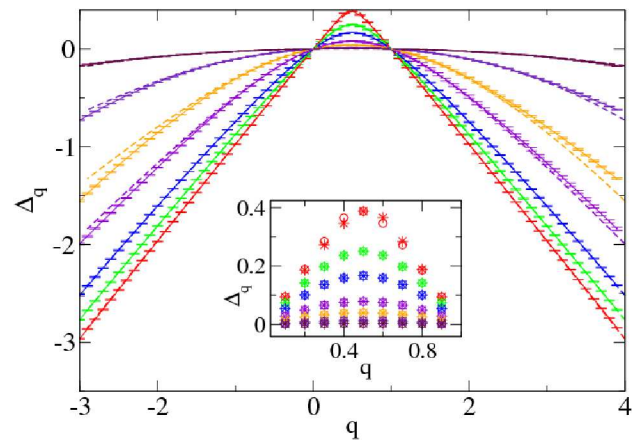}
\end{center}
\caption{(Color online) Anomalous dimensions $\Delta_q$ (solid) and $\Delta_{1-q}$ (dashed) as a function of $q$ for CM$_t$ ensemble, same data and same color code as in Fig.~\ref{fig4}. $g$ increases from bottom curve to top curve at $q=4$. Inset: same data zoomed in, circles correspond to $\Delta_q$, stars to $\Delta_{1-q}$. \label{fig5}}
\end{figure}

\begin{figure}[ht]
\begin{center}
\includegraphics[width=.95\linewidth]{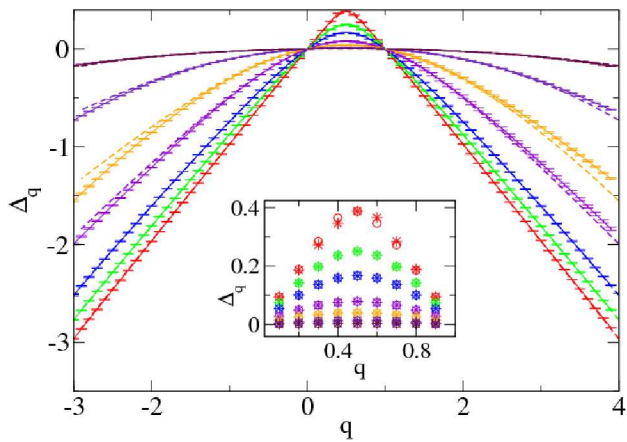}
\end{center}
\caption{(Color online) Fractal dimensions $D_q$ as a function of $q$ for RS ensemble for (from bottom to top at $q=4$) $a=0.01$ (black), 0.05 (red), 0.1 (green), 0.2 (blue), 0.3 (violet), 0.5 (brown), 0.7 (maroon), 0.9 (cyan).  Matrix sizes for numerical fit are $N=2^n$, $8\leq n\leq 12$. Average is performed over all eigenvectors. Number of random realizations of the matrix is between 1024 for $N=2^8$ and 64 for $2^{12}$. Dashed curve is the function $q\mapsto (2q-1)/(q-1)$. \label{fig6}}
\end{figure}

\begin{figure}[ht]
\begin{center}
\includegraphics[width=.95\linewidth]{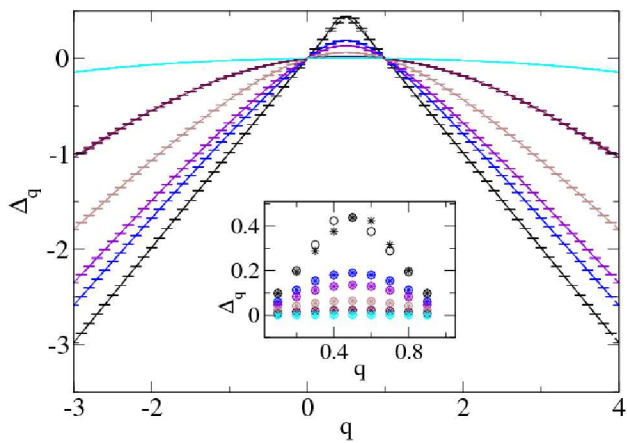}
\end{center}
\caption{(Color online) Anomalous dimensions $\Delta_q$ (solid) and $\Delta_{1-q}$ (dashed) as a function of $q$ for RS ensemble, same data and same color code as in Fig.~\ref{fig6}. $g$ increases from bottom curve to top curve on the right; for better clarity, data corresponding to $a=0.05$ and $0.1$ have been removed. For RS ensemble, the plot for $\Delta_{1-q}$ is almost indistinguishable from the plot for $\Delta_q$. Inset: same data zoomed in, circles correspond to $\Delta_q$, stars to $\Delta_{1-q}$. \label{fig7}}
\end{figure}

\begin{figure}[ht]
\begin{center}
\includegraphics[width=.95\linewidth]{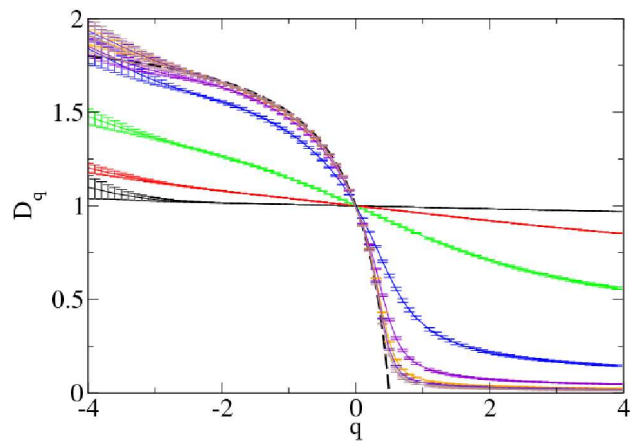}
\end{center}
\caption{(Color online) Fractal dimensions $D_q$ as a function of $q$ for CrBRME with $\beta=2$. Constant is (from top to bottom at $q=4$) $1/g=0.1$ (black), 0.5 (red), 2 (green), 10 (blue), 30 (violet), 60 (orange), 80 (indigo), 100 (brown).  Matrix sizes for numerical fit are $N=2^n$, $8\leq n\leq 12$. Average is performed over the $N/16$ eigenvectors closest to the eigenvalue $E=0$. Number of random realizations of the matrix is between 2560 for $N=2^8$ and 80 for $2^{12}$. Dashed curve is the function $q\mapsto (2q-1)/(q-1)$. \label{fig8}}
\end{figure}

\begin{figure}[ht]
\begin{center}
\includegraphics[width=.95\linewidth]{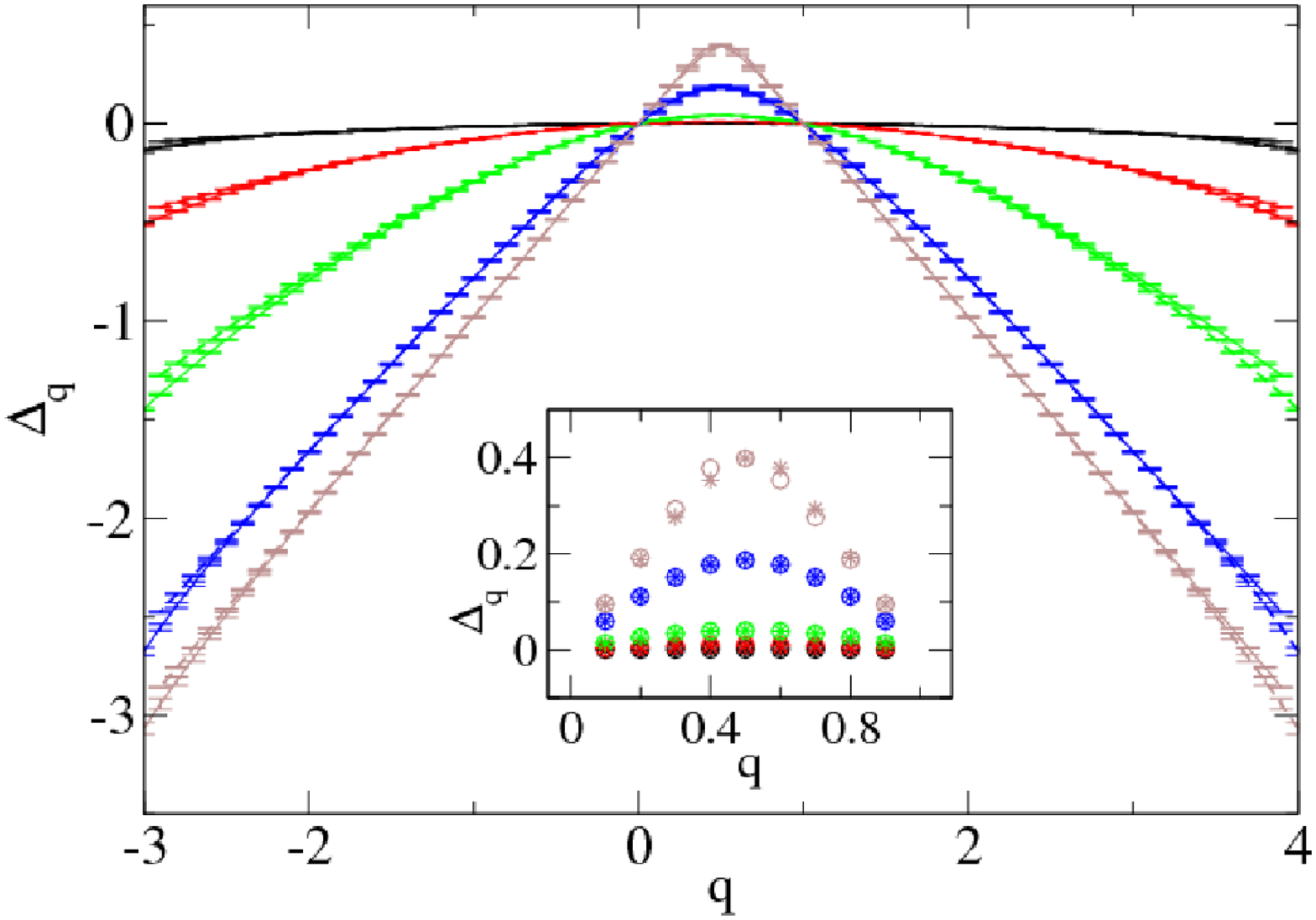}
\end{center}
\caption{(Color online) Anomalous dimensions $\Delta_q$ (solid) and $\Delta_{1-q}$ (dashed) as a function of $q$ for CrBRME, same data as in Fig.~\ref{fig8}. ($g$ increases from bottom curve to top curve on the right; for better clarity, data corresponding to  $1/g=30$, 60 and 80 have been removed).  Inset: same data zoomed in, circles correspond to $\Delta_q$, stars to $\Delta_{1-q}$.\label{fig9}}
\end{figure}

Figures \ref{fig3}, \ref{fig5}, \ref{fig7} and \ref{fig9} are a direct check of the symmetry of anomalous dimensions $\Delta_q$ for ensembles CM$_r$, CM$_t$, RS, and CrBRM, respectively. As expected from the results obtained above, the symmetry is well verified in both limits of small and large coupling constant (or in the case where $a$ is close to an integer for RS ensemble). Our figures show that the symmetry is also quite accurately verified for all other values of the coupling constant, both for $q\in [0,1]$ (see insets of figures \ref{fig3}--\ref{fig9}), and for larger values of $q$.

\section{Intermediate map} \label{sec_interm}
\begin{figure}[t]
\begin{center}
\includegraphics[width=.95\linewidth]{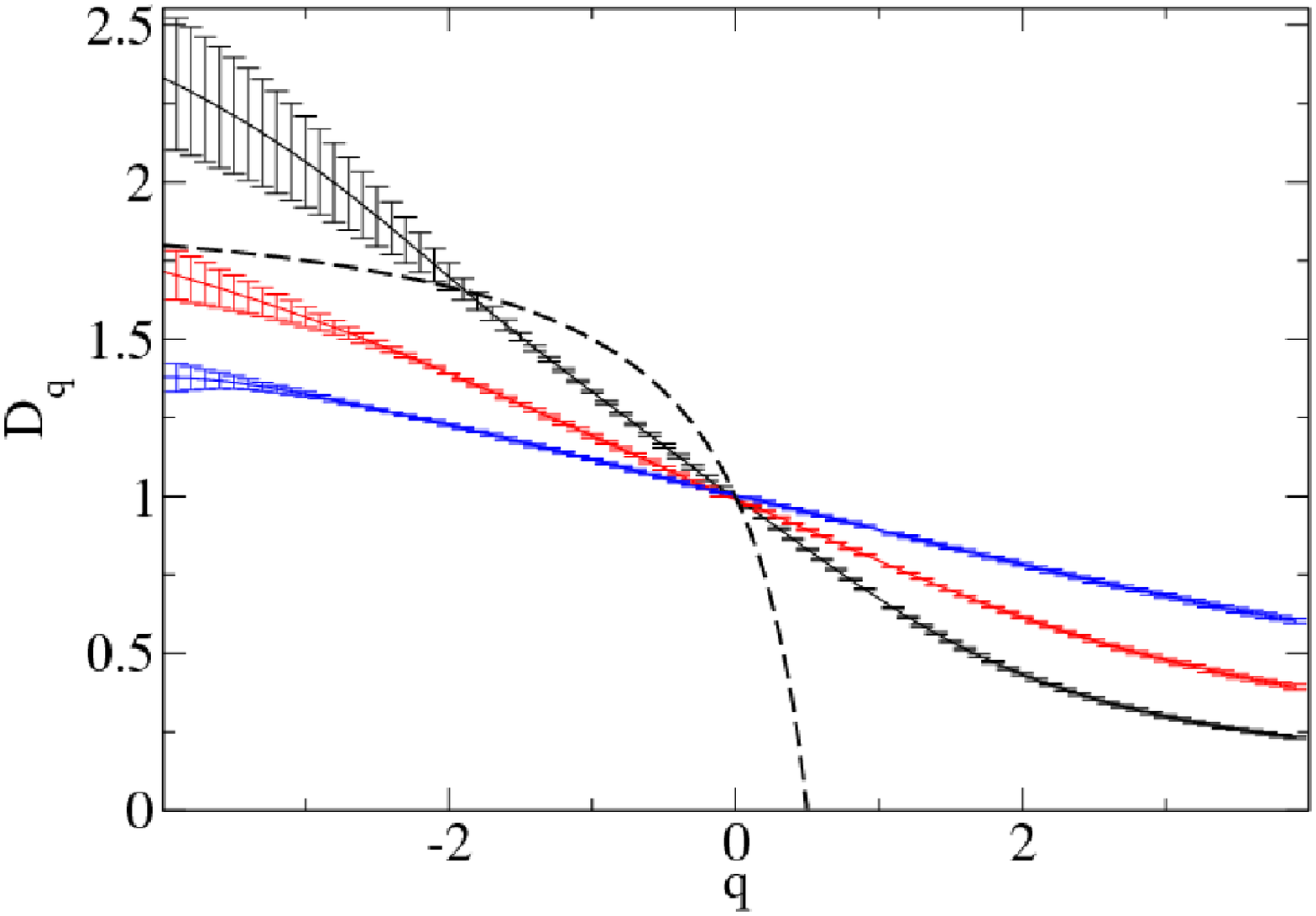}
\end{center}
\caption{(Color online) Fractal dimensions $D_q$ as a function of $q$ for intermediate map for $a=1/b$ with (from bottom to top on the right) $b=3$ (black), 5 (red), 9 (blue).  Matrix sizes for numerical fit are the smallest integer $N'\geq N=2^n$, $8\leq n\leq 12$ such that $N'\equiv 1\ [b]$. Average is performed over all eigenvectors. Number of random realizations of the matrix is between  1024 for $N=2^8$ and 64 for $2^{12}$. Dashed curve is the function $q\mapsto (2q-1)/(q-1)$. \label{fig10}}
\end{figure}

\begin{figure}[ht]
\begin{center}
\includegraphics[width=.95\linewidth]{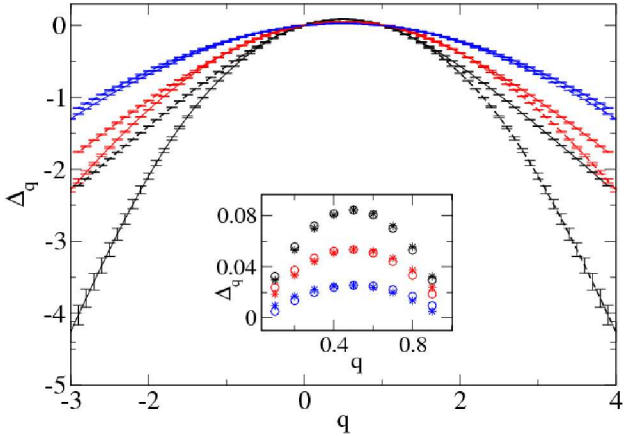}
\end{center}
\caption{(Color online) Anomalous dimensions $\Delta_q$ (solid) and $\Delta_{1-q}$ (dashed) as a function of $q$ for the intermediate ensemble \eqref{interm}, same data as in Fig.~\ref{fig10}. Inset: same data zoomed in.\label{fig11}}
\end{figure}

Although the symmetry of anomalous exponents can be observed in a variety of systems, there are however situations where no symmetry can be observed in the case of strong multifractality. This is the case for instance for the ensemble of $N\times N$ unitary matrices given by 
\begin{equation}
M_{mn}=\mathrm{e}^{\mathrm{i}\Phi_m}\frac{\sin(\pi a N)}{N\sin\left(\pi(m-n+a N\right)/N)},
\label{interm}
\end{equation}
where $\Phi_m$ are independent uniformly distributed random variables, and $a$ some fixed number. This ensemble is identical to RS ensemble, but with a constant $g=aN$ which now depends on the size of the matrix. Spectral statistics of this model turn out to be very different from those of the RS ensemble \cite{BogDubSch}. For rational $a=m/b$ it displays intermediate spectral properties which depend on the remainder of $m N$ modulo $b$. The multifractality spectrum for such an ensemble was investigated in \cite{simplemaps}. The strong multifractality regime is illustrated in Fig.~\ref{fig10} for rational values $a=1/b$ with $b=3,5$ and $9$. As can be seen, its features are quite different from those of other ensembles: for the smallest denominators the multifractal dimensions are quite far from the limiting value $(2q-1)/(q-1)$. Note that in this model there is no clear way to extract a perturbation series expansion. As for the symmetry of anomalous dimensions, it does not hold in general. Indeed, one can see from Fig.~\ref{fig11} that the value of $\Delta_{1-q}$ clearly deviates from  $\Delta_{q}$ for small values of the denominator $b$.

\section{An exactly solvable case}
\label{sec_toy}
We now briefly turn to an example closely related to the above RS ensemble, and where multifractal dimensions can be calculated exactly. Let us consider vectors of the form
\begin{equation}
\label{vectorpsi}
\Psi_p=\frac{\sin(\pi a)}{N\sin(\pi(p+a)/N)}
\end{equation}
for $1\leq p\leq N$, $a$ being a fixed real number.
These vectors correspond to column vectors of the matrix \eqref{RS} for RS ensemble. Namely, $M_{kp}=\mathrm{e}^{\mathrm{i}\Phi_k}\Psi_{k-p}$, and since $M_{kp}$ is unitary, vectors \eqref{vectorpsi} are normalized by
\begin{equation}
\label{vectorpsinormalise}
\sum_{p=0}^{N-1}|\Psi_p|^2=1.
\end{equation}
Our goal in this section is thus to calculate fractal dimensions of a system whose eigenvectors would be columns of the unitary matrix $M_{kp}$. As we now show, multifractal dimensions for the vectors \eqref{vectorpsi} coincide with the zero-order term in the perturbation series expansion of multifractal dimensions for eigenvectors of CM, RS and CrBRM ensembles in the strong multifractality regime, that is, Eqs.~\eqref{dqstrong} and \eqref{dqstrong_sym} when $g=0$.

Moments of the vectors \eqref{vectorpsi} are given by 
\begin{equation}
\mu_{2q}=\sum_{p=0}^{N-1}|\Psi_p|^{2q}=\Big (\frac{\sin (\pi a)}{N}\Big )^{2q} f_N(q,a)\ ,
\label{mu_2q}
\end{equation}
where 
\begin{equation}
\label{fN}
f_N(q,a)=\sum_{p=0}^{N-1}\frac{1}{|\sin(\pi(p+a)/N) |^{2q}}\ .
\end{equation}
In order to get fractal dimensions, we need to calculate the asymptotic behavior of this sum in the limit of large $N$. The usual way of estimating such sums is the Euler-Maclaurin formula. Here we need only the leading term in powers of $N$. 

The sum \eqref{fN} is periodic in $a$ with period $1$ and it is sufficient to consider $0<a<1$. Let $M=[N/2]$ and
\begin{equation}
g_N(q,a)=\sum_{p=0}^{M-1}\frac{1}{\sin^{2q}(\pi(p+a)/N)}\ .
\label{two_parts}
\end{equation}
In the case where $q<1/2$, the sum $g_N(q,a)/N$ can be easily lower and upper bounded by Riemann sums which for large $N$ both converge to the integral
\begin{equation}
\frac{1}{\pi}\int_0^{\pi/2}\frac{\mathrm{d}t}{\sin^{2q}(t)}=\frac{\Gamma(1/2-q)}{2\sqrt{\pi}\Gamma(1-q)}\ .
\end{equation}
For $q>1/2$, using the inequality $x-x^3/6\leq\sin x\leq x$ over $[0,\pi/2]$ one gets  
\begin{eqnarray}
\label{encadrement}
\sum_{p=0}^{M-1}\frac{1}{(p+a)^{2q}}\leq \frac{\pi^{2q}g_N(q,a)}{N^{2q}}\ ,\hspace{4cm}\\
\frac{\pi^{2q}g_N(q,a)}{N^{2q}}\leq \sum_{p=0}^{M-1}\frac{1}{(p+a)^{2q}}\frac{1}{[1-\frac{\pi^2}{6N^2}(p+a)^2]^{2q}}.
\nonumber
\end{eqnarray}
The left-hand side sum tend to $\zeta(2q,a)$ when $N\to\infty$, where $\zeta$ is the Hurwitz zeta function (see e.g.~\cite{bateman}) defined when $s>1$ as 
\begin{equation}
\label{zeta}
\zeta(s,a)=\sum_{p=0}^{\infty}\frac{1}{(p+a)^s}\ .
\end{equation} 
It is not difficult to show that the right-hand side also tends to the same limit when $q>1/2$. Thus from \eqref{encadrement} we obtain that $\pi^{2q}g_N(q,a)/N^{2q}$ converges to $\zeta(2q,a)$. 

Summing up the results we get for $q>1/2$
\begin{equation}
\lim_{N\to \infty}\frac{1}{N^{2q}}f_N(q,a)=\frac{1}{\pi^{2q}}\Big [ \zeta(2q,a)+\zeta(2q,1-a) \Big ]\ .
\label{plus}
\end{equation}
and for $q<1/2$
\begin{equation}
\lim_{N\to \infty}\frac{1}{N}f_N(q,a)=\frac{\Gamma(1/2-q)}{\sqrt{\pi}\, \Gamma(1-q)}\ .
\label{minus}
\end{equation}
For $q=1/2$  one has an additional  $\log N$ term
\begin{equation}
\label{logarithme}
\lim_{N\to \infty} \frac{1}{N} f_N(\frac{1}{2},a)\longrightarrow \frac{2}{\pi}\Big [ \ln N -\ln \frac{\pi}{2}-
\frac{1}{2}(\Psi(a)+\Psi(1-a)) \Big ]
\end{equation} 
where $\Psi(z)$ is the digamma function.

\begin{figure}[t]
\begin{center}
\includegraphics[width=.95\linewidth]{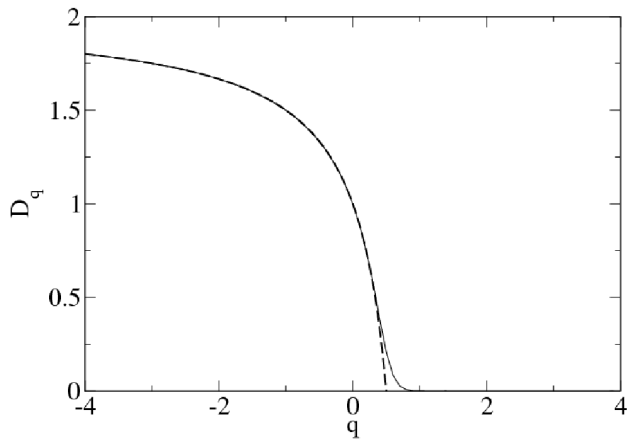}
\end{center}
\caption{(Color online) Fractal dimensions $D_q$ as a function of $q$ for column vectors of Ruijsenaars matrix (vectors \eqref{vectorpsi}) with $a=0.5$, fit over sizes $2^8$ to $2^{12}$. Dashed: theoretical expression $D_q=(2q-1)/(q-1)$ for $q<1/2$, $D_q=0$ for $q>1/2$.  \label{fig12}}
\end{figure}

From these results it follows that the fractal dimensions $D_q$ of vectors \eqref{vectorpsi} is given by
\begin{equation}
D_q=\left \{\begin{array}{cc}0&\mathrm{for}\;q>1/2\\ \displaystyle{\frac{2q-1}{q-1}}&\mathrm{for}\;q<1/2\end{array}\right . \ ,
\label{final_D}
\end{equation}
which exactly obeys the symmetry relation \eqref{symmetry} and coincides with the perturbation series result at $g=0$ (dashed line in Figs.~\ref{fig2}, \ref{fig4} and \ref{fig6}). In Fig.~\ref{fig12} we present the plot of $D_q$ for vectors \eqref{vectorpsi} with $a=.5$. Numerical results agree very well with our analytical result \eqref{final_D} when $q>1$ and $q<0$. However when $q$ gets close to $1/2$, both terms in (\ref{plus}) and (\ref{minus}) contribute for finite values of $N$ and it is difficult to numerically recover Eq.~\eqref{final_D} due to the existence of the logarithmic term in \eqref{logarithme}. 

\section{Conclusion}
We considered critical random matrix ensembles whose elements decrease as inverse powers of their distance to the diagonal. It is known that for such ensembles eigenfunctions have multifractal properties, which are characterized by a nontrivial set of multifractal dimensions $D_q$. Perturbation theory, which is the main analytical tool to access these multifractal dimensions, was successfully applied before, both in the weak multifractality limit (large coupling constant) and in the strong multifractality limit (small coupling constant). However in the latter case, only the region $q>1/2$ had been investigated, since the methods used for this region cannot be directly applied to the region $q<1/2$. 

In this paper we filled this gap by developing a perturbation theory valid in the region $q<1/2$ for small coupling constant. We considered critical random matrix ensembles built from Lax pairs of classical integrable $N$--particle systems, as well as the well-studied critical banded random matrix ensemble. We derived a perturbation expansion of the moments of eigenfunctions
\begin{equation}
\langle\sum_i|\Psi_i|^{2q}\rangle = C_q(g) N^{-D_q(g)(q-1)}
\end{equation}
and obtained the leading terms of the perturbation series in $g$ for multifractal dimensions,
\begin{equation}
D_q(g) \simeq \frac{2q-1}{q-1}+4 |g| \rho(E)\, s\, \frac{ \sqrt{\pi}\, \Gamma \Big (\frac{1}{2}-q\Big)}{(q-1)\Gamma(-q)}\ .
\end{equation}
At order zero, $C_q\sim g^{2q}$ and multifractal dimensions are given by a simple but nontrivial term $D_q=(2q-1)/(q-1)$. The next order is obtained by considering the second-order term of perturbation theory. Higher-order terms can be obtained straightforwardly by the same method.

It turns out that the above formula for $D_q$ at $q<1/2$ exactly coincides with the expression obtained from $D_q$ for $q>1/2$ when assuming that a symmetry holds for anomalous dimensions $\Delta_q=\Delta_{1-q}$, where $\Delta_q=(D_q-1)(q-1)$. Therefore, our results also prove that this conjectured symmetry relation is valid for the models considered at leading orders of perturbation theory in the strong multifractality regime. We also performed detailed  numerical investigations which indicate that this symmetry property is in fact fulfilled for all coupling constants. 

Nevertheless, this symmetry is not an intrinsic property of all critical random matrix ensembles. Indeed, by a small modification of one of our models, we constructed an ensemble where numerics clearly show that such a symmetry is absent. Therefore the range of validity of this symmetry relation remains an open problem (see however the recent paper \cite{mirlin11}).\\

\textit{Acknowledgements} -- The authors are thankful to V.~Kravtsov for many useful discussions and for pointing out Ref.~\cite{mirlin11}. One of the authors (E.~B.~) is indebted to ICTP for hospitality during the visit where part of this work has been done.


\end{document}